# Horizontally oriented plates in clouds

François-Marie Bréon[1], Bérengère Dubrulle[2]

1: Laboratoire des Sciences du Climat et de l'Environnement

CEA/DSM/LSCE, 91191 Gif sur Yvette, France

2: Groupe Instabilité et Turbulence, URA-2464

CEA/DSM/DRECAM/SPEC, 91191 Gif sur Yvette, France

**Abstract**

Horizontally oriented plates in clouds generate a sharp specular reflectance signal in the glint direction, often referred to as "subsun". This signal (amplitude and width) may be used to analyze the relative area fraction of oriented plates in the cloud top layer and their characteristic tilt angle to the horizontal. We make use of spaceborne measurements from the POLDER instrument to provide a statistical analysis of these parameters. More than half of the clouds show a detectable maximum reflectance in the glint direction, although this maximum may be rather faint. The typical effective fraction (area weighted) of oriented plates in clouds lies between $10^{-3}$ and $10^{-2}$. For those oriented plates, the characteristic tilt angle is less than 1 degree in most cases. These low fractions imply that the impact of oriented plates on the cloud albedo is insignificant. The largest proportion of clouds with horizontally oriented plates is found in the range 500-700 hPa, in agreement with typical in situ observation of plates in clouds.

We propose a simple aerodynamic model that accounts for the orienting torque of the flow as the plate falls under its own gravity and the disorienting effects of Brownian motion and atmospheric turbulence. The model indicates that the horizontal plate diameters are in the range 0.1 to a few millimeters. For such sizes, Brownian forces have a negligible impact on the plate orientation. On the other hand, typical levels of atmospheric turbulence lead to tilt angles that are similar to those estimated from the glint observation.

# 1. Introduction

Airborne visual observation of clouds from above sometime shows a glint like-feature that seems to originate from the cloud layer. We have personally observed this pattern on four occasions, all of them from relatively low cloud layers (i.e. 1-3 km) during commercial flight over land surfaces in France and Canada. The narrowness of the glint signature indicates that it is generated by elements that are large compared to the visible light wavelength (little diffraction) and that are perfectly horizontally oriented. For a while, we have been doubtful about this interpretation, thinking that we might be mislead by a surface contribution transmitted through the cloud layer. Indeed, the presence of small water bodies over land surfaces generates very intense and narrow glint patterns, as can be identified during clear sky flights, that are similar to the observation made over the clouds. The question is whether the cloud optical thickness is large enough to eliminate the direct surface contribution to the signal (the surface contribution that is scattered in the cloud cannot generate a highly anisotropic signature).

The question was settled with the observation of a similar very narrow glint pattern from a cloud layer over the ocean. In the vast majority of cases, the ocean surface roughness causes the glint pattern generated at the air-water interface to be much broader than that generated by small water bodies or that observed over clouds. The transmission within the cloud can only generate an even broader feature. This observation rules out the surface contribution hypothesis and we are left with the cloud contribution hypothesis. It provides further evidence that the pattern observed over cloud-covered land is indeed a result of mirror like reflection within the cloud. This interpretation is actually commonly accepted for such observations, referred to as subsuns (Lynch et al. 1994).

In fact, this hypothesis only confirms the interpretation of other optic phenomena, such as the parhelia (Whipple, 1940) and sun pillars (Greenler et al., 1972, Sassen, 1986) that are explained by the presence of horizontally oriented plates in clouds. In situ measurements in clouds confirm the existence of hexagonal plates, and other planar crystals, of various aspect ratios, albeit with no indication on their orientation (Korolev et al., 2000; McFarquhar et al., 2002, Pruppacher and Klett, 1997). Finally lidar measurements pointing to nadir show stronger backscatter signal and lower depolarization ratio than with a few degree off zenith pointing, which is interpreted by the presence of horizontally oriented plates (Platt, 1978; Thomas et al., 1990, Liou et al., 2000, Sassen and Benson, 2001).

Therefore, the presence of horizontally oriented plates in clouds is attested by numerous observations. Chepfer et al (1999) analyze the presence of higher cloud reflectance in the glint direction from spaceborne measurements, which is a strong indication of the presence of horizontally oriented crystal faces in the clouds, and found that this feature is apparent for roughly half of the analyzed clouds. On the other hand, little is known on the fraction of such plates in the cloud, and on the angular spread around the horizontal orientation. In this paper, we make use of specific spaceborne reflectance measurements to quantify the fraction and angular spread of plates with a preferred orientation. We then develop a simple model of plate orientation as a response to aerodynamical torques resulting from the plate fall. The model provides constraints on the plate dimensions that are evaluated in comparison to the observations.

## 2. Measurements

The POLDER (Polarization and Directionality of the Earth Reflectances) instrument was launched onboard ADEOS (ADvanced Earth Observing Satellite) in August 1996. Acquisition was quasi-continuous from late October to the end of June of 1997, when the failure of the platform solar panel terminated the operation of all instruments on board. Nevertheless, eight months of measurements are available and allow original studies thanks to the uniqueness of the measurement principle. Another similar instrument was launched onboard the ADEOS-II platform in December 2002, which itself failed in October 2003. The instrument is composed of a wide field-of-view lens, a filter wheel and a detector (Deschamps et al.; 1994). The filter wheel permits radiance measurements in eight spectral bands from 440 nm (blue) to 910 nm (near IR, water vapor absorption). The detector is a bi-dimensional CCD array with 242 x 274 independent sensitive areas. One snapshot yields an image of a portion of the Earth of size roughly 2400 x 1800 km$^2$, similar to what a camera with a wide field-of-view lens would provide, with a spatial resolution on the order of 6 km equivalent to 0.3°. The pixels in the image are viewed with various zenith angles and azimuths. The zenith angle at the surface varies between 0° at the image center, to 60° crosstrack and 50° forward and aft.

In most cases, depending on the solar position with respect to the satellite, there is one pixel that is observed exactly (at the POLDER pixel angular resolution) in the glint geometry. The pixels surrounding this particular pixel are observed with a slightly different viewing geometry. Assuming that the target does not change significantly between the pixels (details below), this opens the way for a measurement of the reflectance directional signature within a few degrees of the glint direction. The change in glint angle between two adjacent pixels is typically 0.3°. It results mostly from the change in viewing geometry rather than the change in illumination geometry.

One such snapshot is acquired, for each spectral band, every 20 s. There is a large overlap of the areas observed by successive snapshots. As a consequence, the area of interest, which is observed with a geometry close to the specular direction, is also observed within a few minutes from very different directions (see Figure 1). These additional viewing directions allow an estimate of the reflectance directional signature for larger variations of the glint angle, and also a measurement of the reflectance spatial variability when the effects of the directional signature are expected to be small. Thus, one may verify that the reflectance variations observed close to the specular direction are the result of the directional signature, and not a spurious consequence of the surface heterogeneity.

In addition, the POLDER instrument has polarization capabilities that are used here to identify the single scattering contribution to the total cloud reflectance. A cloud field always shows some optical thickness spatial variability that generates significant reflectance heterogeneity. Because the glint directional signature is derived from the apparent spatial structure, this heterogeneity generates some noise on the procedure. On the other hand, the polarized reflectance saturates rapidly as a function of optical thickness. This is because the polarized signal is mostly generated by the single scattering contribution (multiple scattering is non polarized). For a cloud optical thickness larger than about 2, the cloud polarized reflectance is sensitive to the polarized scattering phase function and not to the optical thickness. Thus, the polarized reflectance is better suited than the total reflectance to identify single scattering features such as the specular reflectance over horizontally oriented plates.

Figure 2 shows a cloud field observed with the POLDER instrument. This image was acquired over the North Atlantic (notice the coastlines of Iceland and the southern tip of Greenland on the top of the images). The images are RGB color composites using the three polarized channels of POLDER at 865, 670 and 440 nm. The top image is a "classical" view in total light. Clear areas are dark, with a slight bluish color due to atmospheric scattering. The bottom image is the same in polarized light. Note that the color scale ranges from 0 to 0.1, when it is from 0 to 0.8 for the former (reflectance units). Clear areas appear blue because the light generated by molecular scattering is strongly polarized and larger for the blue (440 nm) channel. The white arc band that extends from Iceland to Greenland corresponds to the cloud-bow and is observed for a scattering angle of roughly 140 degrees. It is a very strong indication of the presence of spherical droplets (Bréon and Goloub, 1998, Riedi et al., 2000). At other directions, the cloud field appears rather dark in polarized light except for the glint direction. This direction is at the center of the concentric circles that indicate the off-glint angles by step of 10 degrees. In the glint direction, a stronger reflection is observed both in total light and in polarized light (zoomed images to the right).

The angular width of the brighter area is on the order of 3 degrees. Let us stress that it is not an artifact of the cloud field structure. A similar bright spot is observed in the glint direction for all POLDER acquisitions over this cloud field, and is not apparent when the same area is seen from a different direction. Over the zoomed images, some colors may be observed on the glint pattern (blue to the North, red to the South). These colors do not have a geophysical origin, but result from the non-simultaneous acquisition of the three channels: a given Earth pixel is observed from slightly different directions. When the directional signature is very strong, as is the case here, the small change in viewing geometry result in a significant change in reflectance, and thus an abnormal spectral (color) signature.

Finally, let us point to the small area towards the bottom part of the images, just below the number "52" in the un-zoomed images. This image is rather dark in total light (top) but is bright in polarized light. This feature results from the ocean glint that is strongly polarized (i.e. the polarization ratio is close to 1). The surface glint is much broader than the cloud glint, a direct consequence of the different tilt angles for the cloud crystal and ocean surface. At the very specular point in fig. 2, the cloud presence hides the surface below, so that the observed signature is generated by the cloud only. At the bottom of the image, there is no cloud field so that one can observe the surface contribution. Because of surface roughness, the surface glint pattern is broad and, as a direct consequence, rather faint. In Figure 2, it is depicted in polarized light only because the dynamic is much larger than in total light.

Although this image shows the best example of a subsun that we have looked at, it is far from exceptional. We now develop a procedure for a quantitative analysis of the feature.

## 3. Radiative Transfer Model

In this section, we propose a simple model of radiative transfer within the cloud. The purpose of the model is to account for the reflectance highly anisotropic pattern observed in the glint direction. Since multiple scattering generates radiance with a smooth directional signature, we only consider the single scattering. The comparison with the measurements must account for the multiple scattering contribution that goes in addition to that quantified here. This contribution may be assumed constant over the small solid angle that we consider.

The cloud is composed of a fraction $\alpha$ of plates oriented close to the horizontal, embedded with either randomly oriented plates and/or other crystal shapes that do not generate glint. We assume that the orientation distribution of near horizontal plates follows a simple Gaussian law: The normalized probability $f$ that a plate is tilted by an angle $\theta_n$ is

$$f \, d\omega_n = \frac{1}{\pi \Theta^2} \exp\left(-\left(\frac{\theta_n}{\Theta}\right)^2\right) d\omega_n \tag{1}$$

where $d\omega_n$ is a solid angle and $\Theta$ is a characteristic tilt angle. The normalization of this function uses an approximation that is valid for small $\Theta$. There is no justification for the choice of a Gaussian function other than its simplicity and the fact that it is physically plausible (no discontinuity in the distribution or its derivative). On the other hand, the comparison with the measurements does not permit to favor any of the various distribution functions that we have tried.

The tilt angle $\theta_n$ is related to the sun and view geometry through:

$$\cos\theta_n = \frac{\cos\theta_s + \cos\theta_v}{2\cos(\gamma/2)} \tag{2}$$

with
$$\cos\gamma = \cos\theta_s \cos\theta_v + \sin\theta_s \sin\theta_v \cos\varphi. \tag{3}$$

where $\theta_s$ and $\theta_v$ are the sun and view zenith angles, $\varphi$ is the relative azimuth, and $\gamma$ is the angle between the sun and view directions. Note that, in the principal plane, $\theta_n = |\theta_s - \theta_v|/2$.

Both the observation pixel and the sun have a finite half width of 0.15 and 0.25° respectively. The corresponding tilt angle variations depend on the geometry and are on the order of ±0.15 degrees. The sunlight is incident to the cloud with the sun zenith angle $\theta_s$. As the light penetrates the cloud, the direct beam is attenuated through particle scattering. Let us consider a slab of the cloud with an effective density of scattering surfaces $dS$ [m²/m²]. A fraction $\alpha$ is horizontally oriented, whereas the rest is random. As the direct beam goes though this slab, a fraction $\alpha\, dS$ is intercepted by horizontal plates, and a fraction $(1-\alpha)/(2\mu_s)\, dS$ is intercepted by the non-oriented particles ($\mu_s$ is the cosine of the sun zenith angle).

The total interception is
$$(\alpha + \frac{1-\alpha}{2\mu_s})dS = K_s\, dS. \qquad (4)$$

Only the near-horizontal plates generate a specular signature. Let us consider the particles oriented within $d\omega_n$. The fraction of incoming light intercepted by these particles is $\alpha\, dS\, f(\theta_n)d\omega_n$. The flux that is specularly reflected by the top face of these particles is:

$$dE = E\, \alpha\, dS\, f(\theta_n)d\omega_n\, F(\mu_s) \qquad (5)$$

where $E$ is the incoming direct beam flux at the top of the slab, and $F$ is the Fresnel reflectance. Accounting for internal refraction and reflection, there is an additional contribution that is almost as large as the first term (Sassen, 1987)[1]. The total flux is therefore twice as much as that given in eq.

---

[1] While a fraction of the light intercepted by the plate is directly reflected by its upper face, the largest part is transmitted. Some of it goes through multiple internal reflection and is eventually transmitted through the upper face. After exiting the plate, this contribution has the exact same direction as the first reflection. If $R$ is the Fresnel reflectance (parallel or perpendicular component), the first reflection is proportional to $R$, whereas the various components resulting from internal reflection are proportional to $(1-R)^2\, R^{2n-1}$. where n is the number of reflections on the bottom face. Summing the components leads to a total reflectance proportional to $2R/(1-R)$, which is very close to $2R$.

(5). The solid angles of the plate normal $d\omega_n$ and that of the corresponding reflected radiance $d\omega_r$ are related through $d\omega_r = 4\,\mu_s\,d\omega_n$. Using this relationship, eq. (5) leads to:

$$dL = \frac{dE}{\mu_v} = \frac{E\,\alpha}{2\mu_s\mu_v} f(\theta_n)\,F(\mu_s)\,dS \tag{6}$$

The double path (down and up) transmittance between the top of the cloud and a level z is

$$T^{\Downarrow\Uparrow} = \exp\left[-\int_0^z (K_s + K_v)\,dS\right] \tag{7}$$

and $E = E_0\,T^{\Downarrow\Uparrow}$ where $E_O$ is the top of the atmosphere incoming irradiance.

The total radiance is the vertical integral of each cloud slab contribution. Assuming a significant optical thickness (larger than about 2 so that the transmission in eq. (7) takes negligible values), integration of equations (6) and (7) leads to:

$$L = \int_0^\infty dL = E_0 \frac{\alpha\,f(\theta_n)F(\mu_s)}{2\mu_s\mu_v(K_s + K_v)} \tag{8}$$

Measurements indicate that $\alpha$ is much smaller than 1. In such case, the expressions for $K_x$ simplify. The reflectance is:

$$R = \frac{\pi L}{E_0} = \frac{\pi\,\alpha\,f(\theta_n)F(\mu_s)}{(\mu_s + \mu_v)} = \frac{\alpha\,F(\mu_s)}{(\mu_s + \mu_v)\,\Theta^2}\exp\left(-\left(\frac{\theta_n}{\Theta}\right)^2\right) \tag{9a}$$

Similarly, the polarized reflectance is:

$$R_P = -\frac{\pi Q}{E_0} = \frac{\pi\,\alpha\,f(\theta_n)F_P(\mu_s)}{(\mu_s + \mu_v)} = \frac{\alpha\,F_P(\mu_s)}{(\mu_s + \mu_v)\,\Theta^2}\exp\left(-\left(\frac{\theta_n}{\Theta}\right)^2\right) \tag{9b}$$

where $Q$ is the second element of the Stokes vector, and $F_p$ is the polarized Fresnel reflectance (Bréon et al., 1995, eq. 5-6).

This expression relates the highly anisotropic reflectance signal generated by horizontally oriented plates to the relative fraction of such plates in the cloud field and to the angular spread of their orientation. The single scattering reflectance measurement probes the cloud top (i.e. the layer with a typical optical thickness of 1). POLDER measurements such as those shown in Figure 2 can be used to retrieve $\alpha$ and $\Theta$.

Within a few degrees of the glint direction, all terms in equation (9) except for the exponential show relatively small variations. Assuming that $\Theta$ is on the order of 1° (0.02 radians) or smaller, the reflectance directional signature is then mostly a function of the tilt angle.

Figure 3 shows an example of POLDER measurements together with the fitted result. The polarized reflectance is shown as a function of the tilt angle $\theta_n$ that was computed from the viewing geometry as in equation (2). The blue, green and red symbols correspond to measurements at 440, 670 and 865 nm, respectively, i.e. the three polarized channels of the instrument. The lines indicate the result of the fit as in equation (9b). No fit is attempted in the blue (440 nm) band because of the significant molecular scattering. The group of symbols for a small range of tilt angles corresponds to a single acquisition shot from the instrument. The spatial variation in the image is here interpreted as a directional feature. On the other hand, the various clusters are the result of POLDER multidirectional acquisition (four are shown in the figure when up to fourteen are available). One may verify that the large directional signal close to zero tilt angle is not apparent for other tilt angles, which is a strong evidence that it is not a consequence of spatial heterogeneity. On this particular case, the retrieved $\alpha$ and $\Theta$ are $7\ 10^{-3}$ and 0.4 degrees. Note that the uncertainty on the tilt angles, resulting from the finite sun width and observation resolution, yields an uncertainty on $\Theta$ that is on the order of 0.1 degree or less.

## 4. Statistical analysis

Eight months of POLDER data were available at the time of the study (i.e. before the launch of ADEOS-2). From the full POLDER dataset, we have extracted all pixels viewed with a direction close to the glint direction. Cloudy pixels were selected with a simple threshold on the reflectance (670 nm reflectance greater than 0.5 in all directions). This simple test selects snow covered surfaces which are latter discriminated from their apparent pressure (see below). For each cluster of pixels, the measured polarized reflectance was fitted by a simple function of $\theta_n$ with parameters $\alpha$, $\Theta$ and an additive low order polynomial that accounts for other processes that generate polarized reflectance, in particular molecular and cloud particle scattering. The RMS difference between the measurements and the best fit quantifies its quality. We define a signal-to-noise as the ratio of the RMS to the Gaussian function amplitude.

The inversion is done independently for 670 nm and 865 nm. Thus, the comparison of the two estimates provides some indication on the retrieval error. Fig 4 shows a scatter plots of the retrieved $\alpha$ for the two bands. The results are fairly consistent, showing a plate fraction in the range 0.01 to 10%, i.e. a rather large dynamic range. The statistics show that, on average, the two band estimates are within a factor of two. There is less relative variability for the characteristic angle $\Theta$ (Fig. 5). Most retrievals are between 0.4 and 1.5 degrees. Although the two estimates show a fair coherence, the scatter is large in regards to the rather low variability. Yet, the noise in the estimate is less than a few tenths of a degree in most cases.

Figure 6 shows a map of the retrieved plate fractions $\alpha$. Note that no satisfactory cloud fields are found in some places, in particular in the tropical regions (no homogeneous cloud fields with a large enough reflectance). On the other hand, at mid-latitude, many estimates are available so that we show the geometric mean of the estimates over 2x2° boxes. The main purpose of this map is to demonstrate that there is no apparent land-sea bias, which provides further evidence that the observed feature is not a surface effect.

Figure 7 shows a cumulative histogram of retrieved $\alpha$ for various cloud pressure ranges. The cloud pressure is estimated from the ratio of two reflectance measurements centered on the oxygen A band at 765 nm (Vanbauce et al., 1998). A reflectance peak in the glint direction is observed in a fair number of cases. Roughly half of the cloud fields that are suitable for our analysis (homogeneity and reflectance criteria) present a significant glint-like signature, in agreement with Chepfer et al. (1999). On the other hand, this signature is rather faint. The retrievals indicate a typical fraction of oriented plates in the range 0.1 to 1 percent. Note that, because the characteristic angle is rather small, even such small fractions yield a reflectance signature of several percent (see eq. 9). The cumulative histograms indicate some differences between the various pressure ranges.

Highest clouds (above 500 hPa) have a lower proportion of horizontal plates than lower clouds. The largest proportion of clouds with horizontally oriented plates is found in the range 500-700 hPa.

Figure 8 is based on the same dataset of retrieved $\alpha$. The cumulative histograms are shown for several ranges of latitude. A comparatively small fraction of clouds with oriented plates are observed in the equatorial band. Since cloud fields in the equatorial band result from deep convection, this result is consistent with that concerning high clouds.

Figure 9 is the same as Figure 7 but for the characteristic tilt angle $\Theta$. Only those inversions with a significant signal to noise ratio are used for this figure. For all pressure ranges, most $\Theta$ are found close to 1 degree. Note that the largest tilt angles (i.e. larger than 3 degrees) are almost entirely found in the low atmospheric layers (i.e. below 700 hPa). As for the low range of tilt angles, virtually none of the clouds above 300 hPa show a characteristic angle smaller than 0.5 degree and the relative fraction increases as the pressure increases. These results are discussed in the final section after an analysis of the plate fall aerodynamic.

## 5. Plate fall hydrodynamic model

In this section, we analyze the orientation of a plate that falls in a fluid as a result of its own weight. We assume that the crystal shapes are hexagonal plates. Theoretical and experimental work show that the stability of the plate orientation depends on the Reynolds number $Re = ud/\nu$ where $d$ is the plate diameter, $u$ is the fall speed and $\nu$ is the fluid viscosity. At very low Reynolds number (Re<0.39), viscous processes dominate the flow with no stabilizing torque, and the plate falls keeping its initial orientation (Schmiedel, 1928, Willmarth et al, 1964). For larger Re, 0.39<Re<80, the fluid around the body becomes increasingly turbulent, generating stable rear eddies, in turn resulting in a dynamical torque. The torque tends to orient the plate horizontally (Mahadevan, 1996). In this regime, the plates have a stable horizontal orientation, and the dynamical torque corrects any deviation of this orientation. For even larger Reynolds, Re>80, the fluid around the plate is fully turbulent, leading to intermittent eddy detachment and shedding. In that regime, the motion during the fall depends on the *Froude Number Fr*, either side-to-side oscillation "flutter" (*Fr*<0.67, see Belmonte et al. 1998) or end-over rotation "tumbling" (*Fr*>0.67).

The observation of horizontally oriented plates indicates that the Reynolds number is in the range 0.39<Re<80. For this particular regime, we now seek the deviation from the horizontal as a function of the plate size. It requires the knowledge of the fall speed $u$.

By definition of the drag coefficient $C_D$, the fall speed is related to the plate mass $m$ through:

$$\frac{C_D}{2} \rho_f \frac{\pi}{4} d^2 u^2 = m g \qquad (10)$$

where $\rho_f$ is the fluid density and $m$ is the plate mass ($m = \rho_s h d^2 \pi / 4$, $h$ the plate thickness). The drag coefficient tends to $32/\pi \, \text{Re}^{-1}$ for very small Reynolds number (Oseen, 1927), and to a constant for large ones (Willmarth et al, 1964). A fit through theoretical and experimental data yields:

$$C_D = \frac{32}{\pi} \frac{1}{\text{Re}} \left( 1 + \frac{\text{Re}}{\pi} \right) \qquad (11)$$

From eq. (10) and (11) and the definition of Re, one gets:

$$u^2 + \pi \frac{\nu}{d} u - \frac{\pi^2}{16} \frac{\rho_s}{\rho_f} h g = 0 \qquad (12)$$

where $\rho_s$ is the plate density. The solution of eq. (12) is:

$$u = \frac{\text{Re} \, \nu}{d} = \frac{\pi \nu}{2 d} \left( -1 + \sqrt{1 + \frac{\rho_s}{\rho_f} \frac{h \, g \, d^3}{d \, 4 \, \nu^2}} \right) \qquad (13)$$

As stated above, oriented plates are expected for Reynolds numbers between 0.3 and 80. From equation (13), this range corresponds to plate diameters from about 0.1 to a few millimeters (see Fig. 10). Such sizes are consistent with in situ measurements of the largest crystal sizes in ice clouds. They are very much larger than the solar wavelength used here, which justifies the neglect of diffraction in the radiative transfer model. The terminal velocities predicted by the model for the millimeter size particles are 0.55 m s$^{-1}$, which is consistent with Heymsfield et al. (2002).

We now consider the orientation of a plate during its fall. Several torques act on a plate that are a function of its fall speed, orientation, and rotation. The net effect of these torques is to orient the plates to the horizontal (for the proper range of Reynolds number). In addition, Brownian motion of the air molecules and air turbulence generate stochastic forces that disrupt the plate orientation. The equation that describes the plate inclination around one axis within the plate writes:

$$I\frac{d^2\theta}{dt^2} = -\gamma \frac{d\theta}{dt} - \beta \sin(2\theta + 2\chi) + M(t) \tag{14}$$

where $I$ is the plate inertial momentum, $\gamma$ expresses the friction, $\beta$ expresses the dynamical torque, $M(t)$ is the Brownian torque, and $\chi$ is the vertical angle of fall. For simplicity, we consider that the ice crystal remains in the regime where $\theta$ is close to zero and the fall is vertical, so that

$$I\frac{d^2\theta}{dt^2} = -\gamma \frac{d\theta}{dt} - 2\beta\,\theta + M(t) \tag{15}$$

The fluid viscosity generates a torque that counters rotation. The friction coefficient $\gamma$ can be computed exactly (Happel and Brenner, 1973) in the low Reynolds number limit although, as will be shown below, its exact expression is not needed here.

The dynamical torque results from the non-symmetrical air flow around the plate as it falls with a tilt angle. The torque amplitude is discussed in Katz (1998). The expression for $\beta$ is:

$$\beta = 2\,C_0\,\rho_f\,d^3\,u^2 \tag{16}$$

where $C_0$ is a coefficient determined from experiments to be on the order of 0.2. Note that this value is in agreement with the theoretical value of π/16 obtained for infinitely thin plates (Tanabe and Kaneko, 1994).

For submicron plates, the main stochastic force $M(t)$ that acts on the particles orientations results from Brownian motion (Katz, 1998). For larger particles, atmospheric turbulence may also be significant. The stochastic torque has the property $\langle M(t)M(t+\Delta t)\rangle = D\delta(\Delta t)$ where δ(t) is the Dirac function and $D$ is the kinetic energy at the considered scale. In our simple model, $D$ is proportional to the sum of Brownian motion energy, $kT$, and the fluid turbulence kinetic energy at the scale of the particles $\rho_f d^3 u_t^2 / 2$ where

$$u_t = \varepsilon^{1/3} d^{1/3} \quad d > \left(\frac{\nu^3}{\varepsilon}\right)^{1/4}$$
$$u_t = d\sqrt{\frac{\varepsilon}{\nu}} \quad d < \left(\frac{\nu^3}{\varepsilon}\right)^{1/4} \tag{17}$$

and $\varepsilon$ is the rate of dissipation of turbulent kinetic energy per unit mass (Klett, 1995). In equation (17) the threshold corresponds to the Kolmogorov scale at which turbulent gradients become regular. The second theorem of fluctuation-dissipation [Kubo, 1966] yields

$$D = \gamma \, (kT + \rho_f d^3 u_t^2 / 2) \tag{18}$$

Equation (14) is a generalized Langevin equation. By virtue of the central limit theorem, the stationary distribution of $\theta$ is Gaussian and

$$\langle \theta^2 \rangle = \frac{D}{2 \beta \gamma} = \frac{1}{8 C_0}\left(\frac{2kT}{\rho_f d^3 u^2} + \frac{u_t^2}{u^2}\right) \tag{19}$$

Note that the angle $\theta$ discussed above is a tilt angle along one direction; say the x-axis direction. The same discussion applies to the tilt with respect to the y-axis. Because the stochastic forcing in the two directions is uncorrelated, the resulting tilt angles are uncorrelated. For small angles, the tilt angle with respect to the zenith $\theta_z$ follows $\theta_z^2 \approx \theta_x^2 + \theta_y^2$. Thus,

$$\langle \theta_z^2 \rangle = \frac{1}{4 C_0}\left(\frac{2kT}{\rho_f d^3 u^2} + \frac{u_t^2}{u^2}\right) \tag{20}$$

Figure 10 shows the predicted quadratic tilt angle $\sqrt{\langle \theta_z^2 \rangle}$ as a function of the particle size $d$. Two cases of cloud pressure and temperature are shown: (400hPa, 220K) and (800hPa, 270K) for three levels of atmospheric turbulence from weak to strong. For the $h/d$ ratio, we made use of the empirical relationship found by Heymsfield (1972) for Cirrus cloud plates:

$$h/d = 2.01 \, d^{-0.551} [\mu m] \tag{21}$$

where d is given in $\mu$m.

On the same figure are shown the bounds for the intermediate range of Reynolds Number (0.39-80) that is necessary for a preferred planar orientation during plate fall. The model indicates that stable fall is expected for plates that are larger than 100 $\mu$m and smaller than a few millimeters. A similar conclusion was obtained by Sassen (1980). Such sizes are consistent with in-situ measurements of the largest crystal sizes in ice clouds. For such diameters, our model assumes *d/h* ratios between 6 and 40, which is the proper order of magnitude. The model predicts that, for a given atmospheric turbulence, the tilt angle varies by a factor of two to three over the range of favorable plate sizes. The predicted tilt angle varies by an order of magnitude from weak to strong level of atmospheric turbulence. Typical values for moderate level of turbulence ($\varepsilon=10^{-2}$ $m^2s^{-3}$) are on the order of one degree, i.e. in good agreement with the observation. Such "typical" value of turbulence have been measured at larger scales than those considered here. It is possible that small-scale processes, such as the wake of cloud particle fall, generate additional turbulence at such scales. However, such additional source is not necessary to explain the observations. Note also that Brownian motion only (i.e. without turbulence) predicts much smaller values of tilt angles. For the favorable range of plate sizes, Brownian stochastic torques can be neglected.

## 6. Discussion and Conclusions

The occurrence of glint like patterns over cloud fields is a frequent phenomenon, discernable from the cloud polarized reflectance in more than half of the cases. Let us stress that this observation is possible in polarized light only. In "natural" light, the spatial structure of the cloud reflectance is much larger than in polarized light, which makes it generally impossible to extract the single scattering contribution from the total cloud reflectance. Because the glint pattern is generated by single scattering within the cloud, it was possible to design a very simple model that reproduces the glint cloud reflectance pattern as a function of the effective fraction of horizontally oriented plates within the cloud and of their typical deviation from horizontality. This was possible by making a number of hypotheses. Diffraction was neglected in our model. This may be justified from the fact that only those plates that are larger than typically 100 $\mu$m may orient properly. Their size is then much larger than the solar wavelengths, in which case the geometric optics is appropriate. We assumed a gaussian tilt function for the plates. Although other functions would be possible, the results on the typical tilt and the fraction of plates would not vary significantly from a different choice. The scatter in the data was found too large to favor a function rather than another. Another difficulty in the data processing results from cases where the glint signature is actually so intense that the measurements at its center are saturated. In such case, the inversion is constrained by the wings of the glint signature with additional uncertainty on the inverted parameters. Finally, note that our measurements are sensitive to the cloud top only (a few optical thickness units). The contribution from lower layers is scattered in the upper cloud layers, which effectively wipes out the glint signature.

With these hypotheses, the measurements indicate that the effective fraction of horizontally oriented plates in the cloud is generally in the range 0.1 to 1%. Although glint signatures are observed for clouds at all pressure levels, the largest fractions are found for those clouds in the range 500-700 hPa. At mid-latitudes, where most of the cloud glint signatures are observed (Figure 6), this corresponds to typical temperature on the order of –10 to –30°C, This range of temperature is consistent with that derived from ground-based lidar observation of oriented plates in clouds (Sassen and Benson, 2001).

Although the specular contribution to the reflectance in the glint direction can be larger than the non-specular contribution, it is significant only over a rather small solid angle. The quantitative model indicates that the detected signal is generated by a very small (typically 0.1 to 1%) fraction of the ice crystals (weighted by their effective surface). The horizontal orientation may impact the cloud albedo through a change of the sunlight interception (Takano and Liou, 1989) but not more than a factor of 2. To the first order, the impact scales with the fraction of such horizontally oriented plates. Our results indicate that, although the effect of the horizontally oriented plates on the glint directional signature (and on other optical phenomena) is large, the impact on the cloud albedo is generally insignificant.

The measurements indicate tilt angles that are generally very close to one degree. Such value is fully consistent with the prediction of the aerodynamic model for moderate level of turbulence. The model predicts that the typical tilt angles vary between 0.5 and a few degrees from weak to strong turbulence, a range that is in good agreement with the observations. We also note that the largest tilt angles (i.e. a few degrees) are mostly measured for low clouds ($P_{app}$>700 hPa) where strong levels of atmospheric turbulence are expected. These observations are strong indications of the simple model validity.

Nevertheless, a strong limitation of the aerodynamic model is the assumption that horizontal surfaces are found over hexagonal plates only. In practice, both airborne measurements (e.g. Korolev et al., 2000) and laboratory experiments (e.g. Bailey and Hallet, 2002) indicate that there are many other planar crystals (Pruppacher and Klett, 1997). In particular for the large sizes that appear suitable for oriented fall, there is a tendency for branching to occur. Clearly, for crystal shapes other than the simple hexagonal plate that was considered here, many parameters of the model may change: Effective density is most probably lower, resulting in smaller fall velocities for a given diameter; in addition, both the aerodynamic torque and the moment of inertia depends on the particle shape. As a consequence, the quantitative predictions of our simple model only apply to the hexagonal plates, and are increasingly inaccurate for more complex shapes. On the other hand, as branching increases, some loss of symmetry is expected, in which case the orientation is unlikely to be strictly horizontal. Thus, we hypothesized that surfaces that are horizontal within a few tenths of a degree are only possible for the simplest, highly symmetrical, crystals, i.e. hexagonal plates.

# 7. Acknowledgments

The data in this paper were based on measurements acquired by the CNES/POLDER-1 instrument onboard the NASDA/ADEOS-1 platform. We thank Andy Heymsfield and two anonymous reviewers for valuable comments on an earlier version of the manuscript.

## 9. Figure Captions

**Figure 1**: Typical viewing geometry for a set of 7x7 POLDER pixels used to study the glint directional signature. The horizontal line is the principal plane (that contains the sun direction and local nadir). The full circles indicate the view zenith angle $\theta_v$ by steps of 20°. The ellipses show the glint angle $\xi$ by steps of 10°. The 13 clusters of dots correspond to 13 successive acquisitions within roughly two minutes. Each cluster is composed of 7x7 dots corresponding to 49 contiguous surface pixels of size $(6.2 \text{ km})^2$.

**Figure 2:** POLDER false color image (865, 670 and 440 channels are represented by red, green and blue respectively) showing a sharp reflectance maximum at the specular direction. White regions (glint, cloud bow) result from the overlapping of the 3 primary colors. The green straight line is the principal plane. The red lines indicate the off-glint angles by step of 10 degrees. The figures to the right are a zoom of those to the left. Top images correspond to total reflectance, whereas the bottom ones are images in polarized reflectance.

**Figure 3:** Measured reflectance as a function of the tilt angle. The tilt angle is the angle of the plate necessary to generate a specular reflection to the viewing direction. Red, green and blue crosses correspond to measurements acquired at 865, 670 and 440 nm respectively. The red and green lines show the fit as in equation (9).

**Figure 3 :** Measured reflectance as a function of the tilt angle. The tilt angle is the angle of the plate necessary to generate a specular reflection to the viewing direction. Red, green and blue crosses correspond to measurements acquired at 865, 670 and 440 nm respectively. The red and green lines show the fit as in equation (9).

**Figure 4:** Scatter plot of the retrieved plate fraction $\alpha$ in the two bands. The measurements and the fit are independent. Thus, the scatter provides an indication of the measurement noise.

**Figure 5:** Same as Figure 4 but for the characteristic angle $\Theta$.

**Figure 6:** Global map of the retrieved amplitude. The scale is the decimal logarithm of the plate fraction $\alpha$.

**Figure 7:** Cumulative histogram of the retrieved oriented plate fraction $\alpha$. The cumulative histograms are shown for various apparent cloud top pressures, derived from the ratio of two POLDER measurements centered on the oxygen A band at 763 nm.

**Figure 8:** Same as Figure 7 but the cumulative histograms are shown for various Latitude bands.

**Figure 9:** Same as Figure 7 but for the characteristic tilt angle $\Theta$.

**Figure 10:** Result of the aerodynamical fall model. The quadratic tilt angle $\sqrt{\langle \theta_z^2 \rangle}$ (see equation 20) is shown as a function of the plate diameter. The vertical lines indicate the range of plate diameter for which a stable fall is expected (from the Reynolds number, see equation 13). Plain and dashed lines correspond to two sets of (Pressure, Temperature), i.e. (800 hPa, 270 K) and (400 hPa, 220 K) respectively. The three upper groups of lines correspond to weak ($\varepsilon=10^{-3}$ m$^2$s$^{-3}$) moderate ($\varepsilon=10^{-2}$) and strong ($\varepsilon=10^{-1}$) turbulence from bottom to top. The two bottom lines indicate the tilt angle when only Brownian motion is considered.

## 10. Figures and Captions

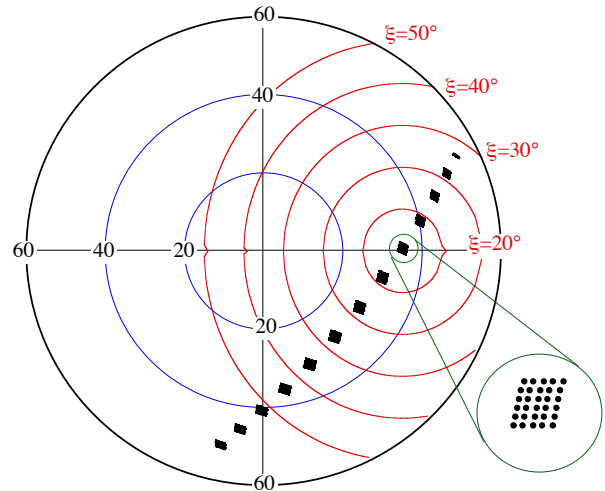

**Figure 1**: Typical viewing geometry for a set of 7x7 POLDER pixels used to study the glint directional signature. The horizontal line is the principal plane (that contains the sun direction and local nadir). The full circles indicate the view zenith angle $\theta_v$ by steps of 20°. The ellipses show the glint angle $\xi$ by steps of 10°. The 13 clusters of dots correspond to 13 successive acquisitions within roughly two minutes. Each cluster is composed of 7x7 dots corresponding to 49 contiguous surface pixels of size $(6.2 \text{ km})^2$.

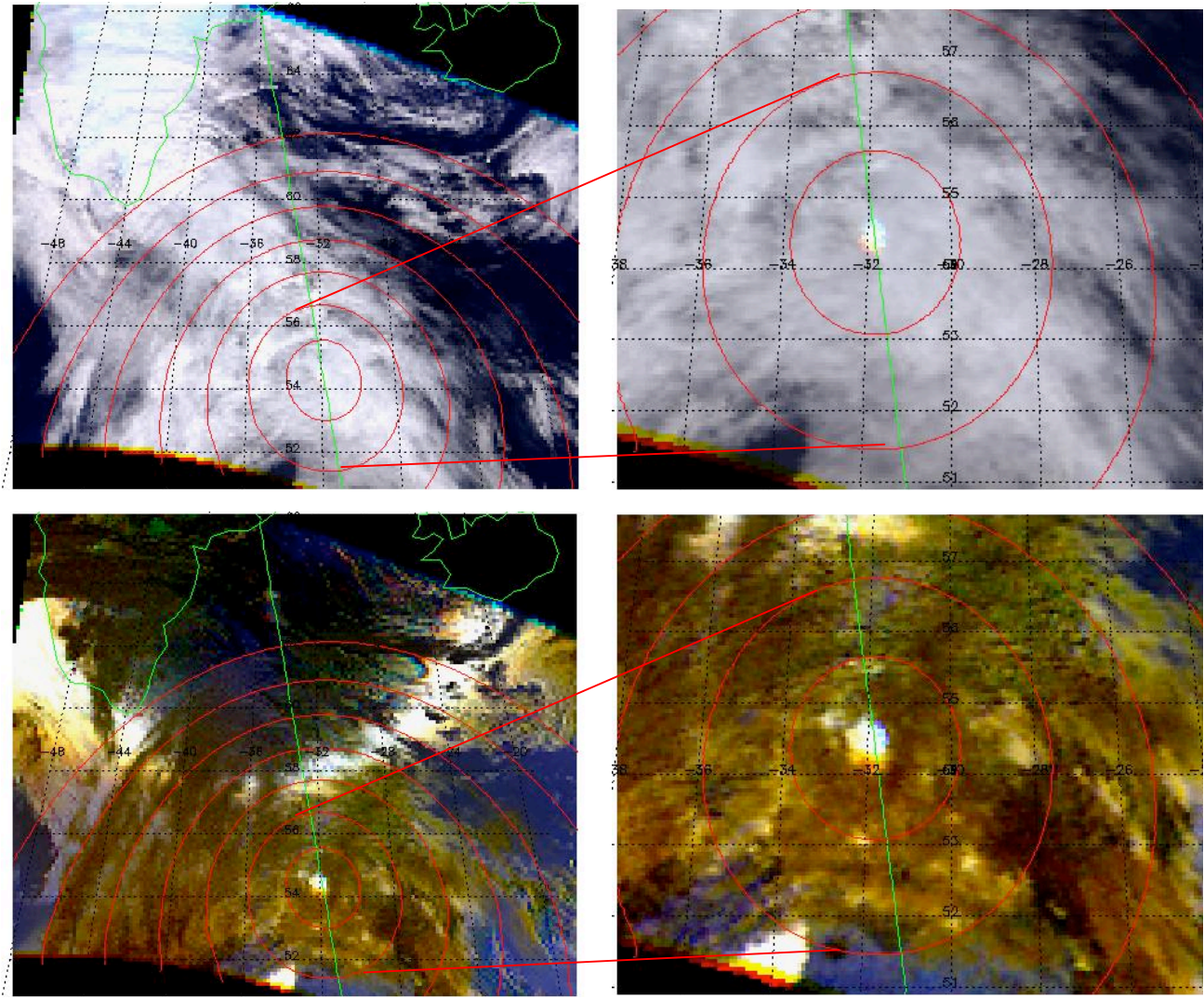

**Figure 2:** POLDER false color image (865, 670 and 440 channels are represented by red, green and blue respectively) showing a sharp reflectance maximum at the specular direction. White regions (glint, cloud bow) result from the overlapping of the 3 primary colors. The green straight line is the principal plane. The red lines indicate the off-glint angles by step of 10 degrees. The figures to the right are a zoom of those to the left. Top images correspond to total reflectance, whereas the bottom ones are images in polarized reflectance.

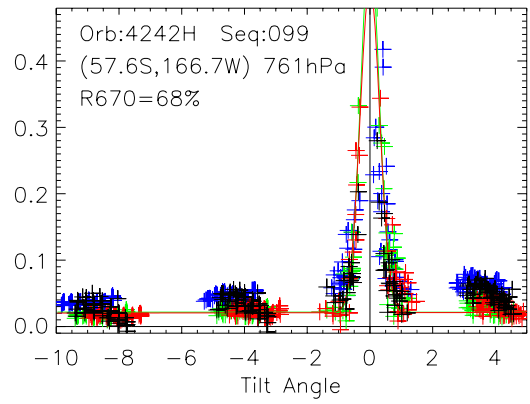

**Figure 3 :** Measured reflectance as a function of the tilt angle. The tilt angle is the angle of the plate necessary to generate a specular reflection to the viewing direction. Red, green and blue crosses correspond to measurements acquired at 865, 670 and 440 nm respectively. The red and green lines show the fit as in equation (9).

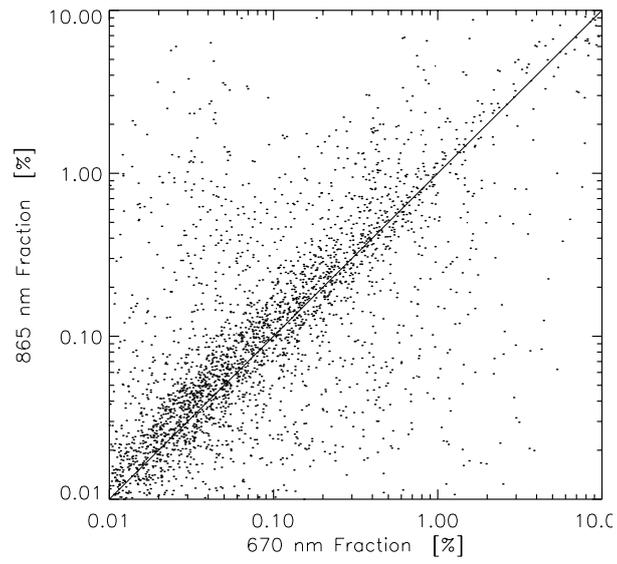

**Figure 4:** Scatter plot of the retrieved plate fraction $\alpha$ in the two bands. The measurements and the fit are independent. Thus, the scatter provides an indication of the measurement noise.

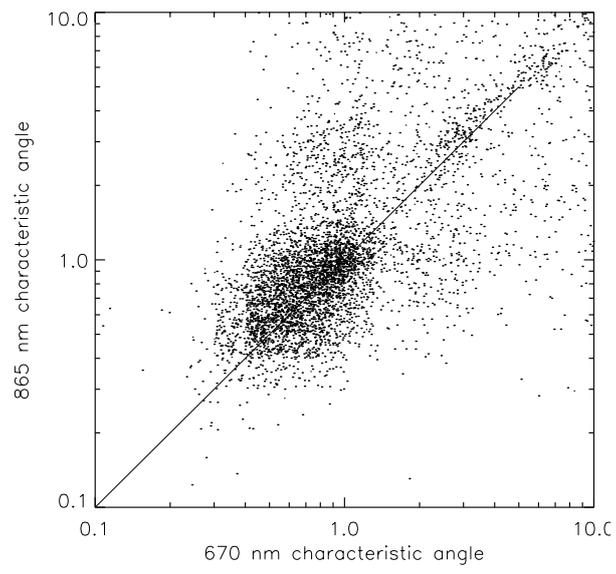

**Figure 5:** Same as Figure 4 but for the characteristic angle $\Theta$.

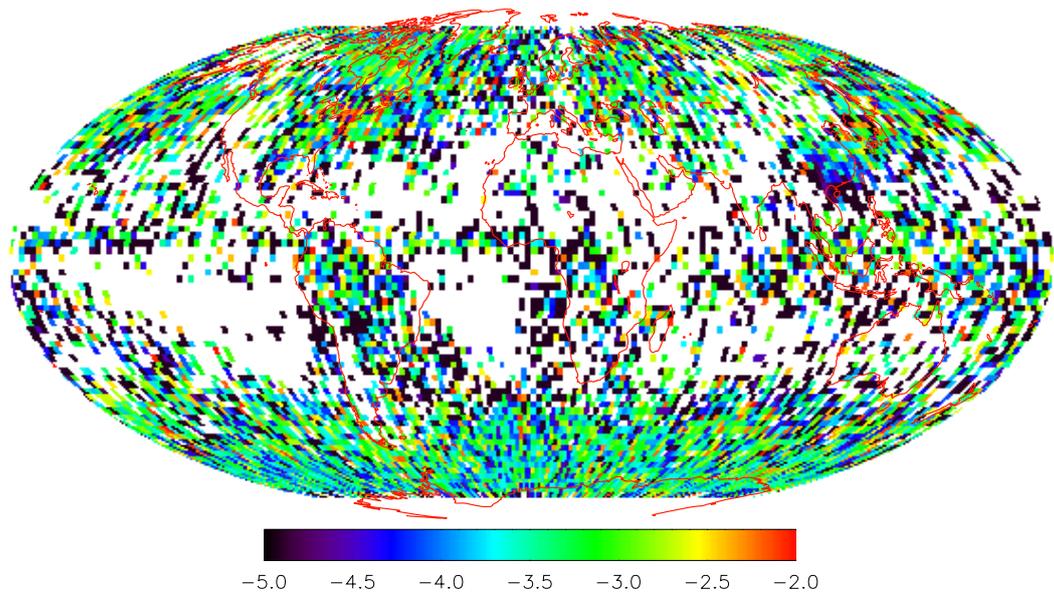

**Figure 6:** Global map of the retrieved amplitude. The scale is the decimal logarithm of the plate fraction $\alpha$.

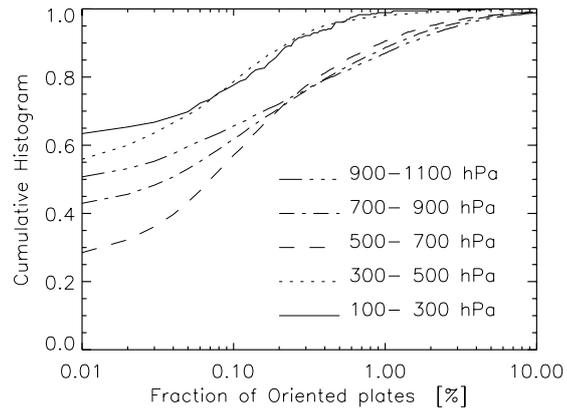

**Figure 7:** Cumulative histogram of the retrieved oriented plate fraction $\alpha$. The cumulative histograms are shown for various apparent cloud top pressures, derived from the ratio of two POLDER measurements centered on the oxygen A band at 763 nm.

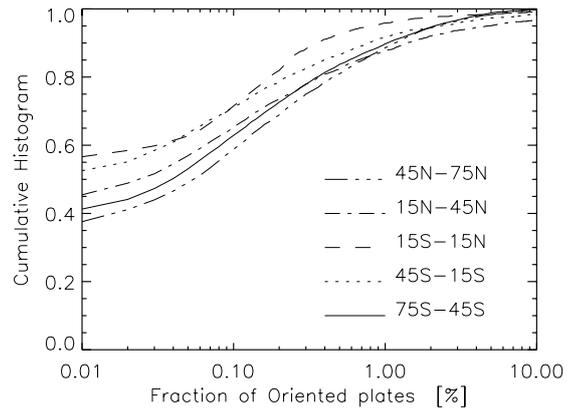

**Figure 8:** Same as Figure 5 but the cumulative histograms are shown for various Latitude bands.

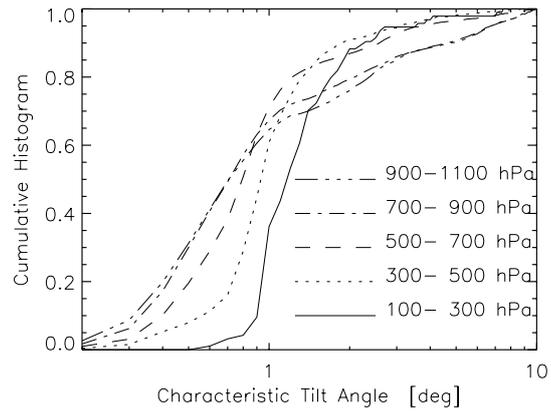

**Figure 9:** Same as Figure 5 but for the characteristic tilt angle *Θ*.

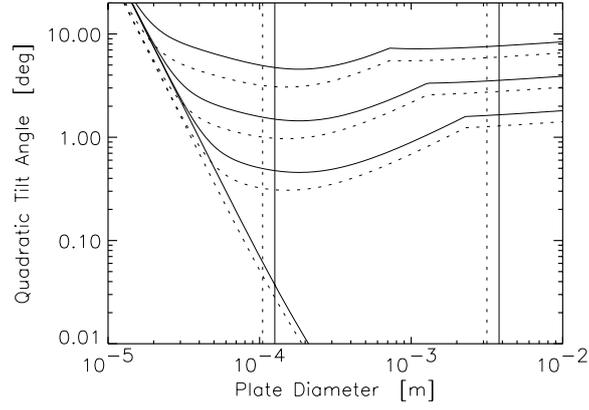

**Figure 10:** Result of the aerodynamical fall model. The quadratic tilt angle $\sqrt{\langle \theta_z^2 \rangle}$ (see equation 20) is shown as a function of the plate diameter. The vertical lines indicate the range of plate diameter for which a stable fall is expected (from the Reynolds number, see equation 13). Plain and dashed lines correspond to two sets of (Pressure, Temperature), i.e. (800 hPa, 270 K) and (400 hPa, 220 K) respectively. The three upper groups of lines correspond to weak ($\varepsilon=10^{-3}$ m$^2$s$^{-3}$) moderate ($\varepsilon=10^{-2}$) and strong ($\varepsilon=10^{-1}$) turbulence from bottom to top. The two bottom lines indicate the tilt angle when only Brownian motion is considered.